\documentclass[12pt,,letterpaper]{JHEP3}
\usepackage{graphics}
\usepackage{epsfig}
\usepackage{slashed}

\title{Holographic non-relativistic fermionic fixed point and bulk dipole coupling}

\author{
Wei-Jia Li\\
Department of Physics, Beijing Normal University, 100875 Beijing,
China\\
\email{wjli@mail.bnu.edu.cn}
}

\author{
Hongbao Zhang\\
Crete Center for Theoretical Physics, Department of Physics, \\
University of Crete, 71003 Heraklion, Greece\\
\email{hzhang@physics.uoc.gr}
}

\abstract{Inspired by the recently discovered non-relativistic fermionic fixed points,
we investigate how the presence of bulk dipole coupling modifies the spectral function
at one of these novel fixed points. As a result, although the infinite flat band is always
visible in the presence of the bulk dipole coupling as well as chemical potential, the band
is modified in a remarkable way at small momenta up to the order of magnitude of bulk dipole coupling.
 On the other hand, like a phoenix, a new Fermi surface sprouts from the formed gap when the bulk dipole
  coupling is pushed up further such as to overshadow the charge parameter, which is obviously different
  from what is found at the relativistic fixed points.}
\preprint{CCTP-2011-34}
\begin{document}
\section{Introduction}
Arising from the string theory, AdS/CFT correspondence has provided
us with a new paradigm to study strongly coupled many body phenomena
by relating them to a single or few body problem in the bulk
classical gravitational backgrounds with one extra dimension. In
applying AdS/CFT correspondence, there are two approaches which have
been taken. One is the so called bottom-up approach, where the bulk
setup is devised in a simple way but the microscopic understanding
of dual field theory is generically lacking. The other is the so
called top-down approach, where the whole setup can be embedded in
the more sophisticated string theory or M theory such that the
microscopic content of boundary field theory is well understood,
although the system may not be realized in Nature. However, such a
shortcoming is somehow mitigated by focusing judiciously on the IR
physics, where it is believed that the same universal behaviors can
be extracted from this sort of holographic duality for those
realistic systems which may have different microscopic contents.

With this in mind, attempts to apply this holographic technique,
with some success, have gone beyond high energy physics,  especially to
condensed matter physics in the recent years. Obviously, it is kind
of win-win game to apply the holography to condensed matter physics.
On the one hand, there exist many strong coupled systems in
condensed matter physics, which are intractable by the conventional
approaches. While AdS/CFT correspondence, as kind of strong/weak
duality, can offer unprecedented insight into the dynamics of these
systems. On the other hand, unlike other disciplines such as high
energy physics and cosmology, only with some tabletop experiments,
does condensed matter physics allow one to cook up matter such that
various vacuum states and phases can be created in the laboratory.
With this available landscape of man-made multiverse to scan, it is
highly possible for us to hit some universality classes extracted from
the holography. In this sense, condensed matter physics may in turn
provide us with the first experimental evidence for AdS/CFT
correspondence. For a review of this exciting subject, please refer
to \cite{Hartnoll,Herzog1,McGreevy,Horowitz,Sachdev}.

In particular, partly triggered by the lack of a proper field
theoretical framework to explain for the mysterious behaviors
 of existing non-Fermi liquids, endeavors have been
 made recently to study the spectral function of fermionic operators by
 holographically manipulating the bulk Dirac field which is minimally coupled to
 gravity and gauge fields, where the emergence of Fermi surface has
 been identified with a rich spectrum of non-Fermi liquid behaviors\cite{Lee,LMV,CZS,FLMV,FILMV,ILM}.
 Later on, inspired by the generic top-down models, the bottom-up
 dipole coupling has been added to the previous minimally coupled Dirac
 field and its effects on the fermionic correlator have been
 investigated, where a possible dynamical gap opens up when
 the dipole coupling is tuned to be large enough\cite{ELP,ELLP,GM}\footnote{It is noteworthy that the impact of bulk dipole coupling
 on fermionic correlator has also been fully explored from the purely top-down construction in \cite{GSW1} and \cite{GSW2}.}. Generically, in the context of AdS/CFT correspondence,
 not only the modification of bulk dynamics, but also the change of boundary conditions can
 alter the boundary field theory. Actually, in the latter manner, the holographic non-relativistic
 fermionic fixed points have been implemented most recently by adding a Lorentz violating
 boundary term instead of the conventional Lorentz covariant one to
 the bulk minimally coupled Dirac action, where a holographic flat band is achieved\cite{LT1}.

 Along this line, this paper is intended to insist on one of these novel non-relativistic fixed points and
 investigate how the fermionic correlator is modified when we turn on the bulk dipole
 coupling. In the next section, we shall provide a brief review of how both of the
relativistic and non-relativistic fermionic fixed points are
implemented by holography. In Section \ref{mainresult}, after
building up the holographic framework to extract the fermionic
correlator at the non-relativistic fermionic fixed point, we shall present our
numerical results for the relevant quantities
associated with the fermionic correlator in the presence of bulk
dipole coupling. Conclusions and discussions will be addressed in
the end.
\section{Holographic implementation of various fermionic fixed points}
Start with the bulk action for a probe Dirac fermion with the mass
$m$, charge $q$ and magnetic dipole coupling $p$
\begin{equation}
S_{bulk}=\int_\mathcal{M}
d^4x\sqrt{-g}i\bar{\psi}\Big[\frac{1}{2}(\overrightarrow{\slashed{D}}-
\overleftarrow{\slashed{D}})-m-ip\slashed{F}\Big]\psi
\end{equation}
in the following fixed background, i.e.,
\begin{equation}\label{blackhole}
ds^2=r^2[-f(r)(dt)^2+(dx^1)^2+(dx^2)^2]+\frac{1}{r^2}\frac{(dr)^2}{f(r)},
A_a=A_t(dt)_a.
\end{equation}
Here $\bar{\psi}=\psi^\dag\Gamma^t$,
$\overrightarrow{\slashed{D}}=(e_\mu)^a\Gamma^\mu[\partial_a+\frac{1}{4}[(\omega_{\rho\sigma})_a\Gamma^{\rho\sigma}-iqA_a]$,
and
$\slashed{F}=\frac{1}{4}\Gamma^{\mu\nu}(e_\mu)^a(e_\nu)^bF_{ab}$,
where $(e_\mu)^a$ form a set of orthogonal normal vector bases, and
Gamma matrices satisfy $\{\Gamma^\mu,\Gamma^\nu\}=2\eta^{\mu\nu}$
with the spin connection
$(\omega_{\mu\nu})_a=(e_\mu)_b\nabla_a(e_\nu)^b$,
$\Gamma^{\mu\nu}=\frac{1}{2}[\Gamma^\mu,\Gamma^\nu]$, and the field
strength $F=dA$. In addition, the emblackening factor and gauge
potential are given by
\begin{equation}
f=1-\frac{1+Q^2}{r^3}+\frac{Q^2}{r^4}, A_t=g_FQ(1-\frac{1}{r}),
\end{equation}
which arises as the charged black hole solution to the equation of motion
following from the bulk action for the gauge field coupled to AdS
gravity with the gauge coupling $g_F$, i.e.,
\begin{equation}
S=\frac{1}{2\kappa^2}\int_\mathcal{M}
d^4x\sqrt{-g}\Big[R-\frac{6}{L^2}-\frac{L^2}{g_F^2}F_{\mu\nu}F^{\mu\nu}\Big],
\end{equation}
where $L$ is the curvature radius, and has been set to unity along
with the horizon radius. By holography, such a charged black hole
places the probe fermion into the dual strongly coupled soup with
the finite temperature and chemical potential as
\begin{equation}
T=\frac{3-Q^2}{4\pi}, \mu=g_FqQ.
\end{equation}

Now to have a well-defined variational principle for the Dirac
action, a boundary term must be added. To see this, let us make the
variation of bulk action, which gives rise to
\begin{eqnarray}
\delta
S_{bulk}&=&i\int_\mathcal{M}d^4x\sqrt{-g}\Big[\delta\bar{\psi}(\overrightarrow{\slashed{D}}-m-ip\slashed{F})\psi-\bar{\psi}(\overleftarrow{\slashed{D}}+m+ip\slashed{F})\delta\psi\Big]\nonumber\\
&&+\frac{i}{2}\int_{\partial\mathcal{M}}d^3x
\sqrt{-h}(\bar{\psi}_-\delta\psi_+-\bar{\psi}_+\delta\psi_-+\delta\bar{\psi}_+\psi_--\delta\bar{\psi}_-\psi_+)
\end{eqnarray}
where $h=\frac{g}{g_{rr}}$ is the determinant of induced metric on
the boundary, and $\psi_\pm=\frac{1}{2}(1\pm\Gamma^r)\psi$. Note
that the bulk Dirac equation
\begin{equation}\label{Dirac}
(\overrightarrow{\slashed{D}}-m-ip\slashed{F})\psi=0
\end{equation}
is first order. So it is not allowable to fix all components of
$\psi$, namely both of $\psi_+$ and $\psi_-$. Instead we must
somehow fix simply half of the components of $\psi$, which can
actually be implemented by adding a boundary term. Moreover, it
turns out that the variational principle can be achieved by adding a
number of different boundary terms, depending on the specific value
of the mass parameter $m$\footnote{For simplicity but without loss
of generality, we shall focus on the case of $m\geq0$ in what
follows.}.

The conventional boundary term chosen to be added is the Lorentz
covariant one, i.e.,
\begin{equation}
S_{bdy}=\frac{i}{2}\int_{\partial\mathcal{M}}\sqrt{-h}\bar{\psi}\psi=\frac{i}{2}\int_{\partial\mathcal{M}}\sqrt{-h}(\bar{\psi}_-\psi_++\bar{\psi}_+\psi_-),
\end{equation}
whereby the variation of the full on-shell action is given by
\begin{equation}
\delta S_{bulk}+\delta
S_{bdy}=i\int_{\partial\mathcal{M}}d^3x\sqrt{-h}(\bar{\psi}_-\delta\psi_++\delta\bar{\psi}_+\psi_-),
\end{equation}
which indeed vanishes if and only if the Dirichlet boundary
condition is imposed on $\psi_+$. This sort of choice of boundary
condition is usually referred to as the standard quantization for the
Dirac field. As a result, the dual boundary field theory is a
Lorentz covariant CFT where loosely speaking $\psi_+$ plays a role
of fermionic source, and the dual operator is given by $\psi_-$ with
dimension
\begin{equation}
\Delta[\psi_-]=\frac{3}{2}+m.
\end{equation}
When the mass parameter is tuned into the window $0\leq
m<\frac{1}{2}$, we can have other boundary conditions to choose. The
first one is the so called alternative quantization, which can be
implemented simply by adding the boundary term with opposite sign,
i.e.,
\begin{equation}
S_{bdy}=-\frac{i}{2}\int_{\partial\mathcal{M}}\sqrt{-h}\bar{\psi}\psi=-\frac{i}{2}\int_{\partial\mathcal{M}}\sqrt{-h}(\bar{\psi}_-\psi_++\bar{\psi}_+\psi_-).
\end{equation}
This now results in a well defined variational principle if and only
if one imposes the Dirichlet boundary condition on $\psi_-$. The
dual boundary field theory is still a Lorentz covariant CFT with the
fermionic source and dual operator interchanged. The dimension of
operator is thus given by
\begin{equation}
\Delta[\psi_+]-=\frac{3}{2}-m.
\end{equation}
We have other boundary conditions to choose if we are not only
interested in the Lorentz covariant boundary field theory. In
particular, as shown in \cite{LT1}, a non-relativistic fermionic fixed
point can be implemented by adding the following Lorentz violating
boundary term, i.e.,
\begin{equation}\label{lvb}
S_{bdy}=\frac{1}{2}\int_{\partial\mathcal
{M}}d^3x\sqrt{-h}\bar{\psi}\Gamma^1\Gamma^2\psi.
\end{equation}
To see explicitly what both of the corresponding fermionic source
and dual operator look like, we would like to firstly choose our
Gamma matrices once and for all as follows
\begin{eqnarray}
\Gamma^r=\left(
  \begin{array}{cc}
    -\sigma^3 & 0 \\
    0 & -\sigma^3 \\
  \end{array}
\right),\ \ \Gamma^t=\left(
             \begin{array}{cc}
               i\sigma^1 & 0 \\
               0 & i\sigma^1 \\
             \end{array}
           \right),\ \ \Gamma^1=\left(
             \begin{array}{cc}
               -\sigma^2 & 0 \\
               0 & \sigma^2 \\
             \end{array}
           \right),\ \ \Gamma^2=\left(
             \begin{array}{cc}
               0 & -i\sigma^2 \\
               i\sigma^2 & 0 \\
             \end{array}
           \right),\nonumber\\
\end{eqnarray}
where $\sigma^i$ are Pauli matrices\footnote{Note that our choice of Gamma matrices is different from that in \cite{LT1}. We are believed that our choice is more suitable for performing the practical calculation, which can be seen later on.}. With this kind of choice of
Gamma matrices, we can express $\psi_+$ and $\psi_-$ as
\begin{eqnarray}
\psi_+=(-h)^{-\frac{1}{4}}\left(
  \begin{array}{c}
    0 \\
    z_1 \\
    0 \\
    z_2 \\
  \end{array}
\right),\ \, \ \psi_-=(-h)^{-\frac{1}{4}}\left(
  \begin{array}{c}
    y_1 \\
    0 \\
    y_2 \\
    0 \\
  \end{array}
\right).
\end{eqnarray}
Hereby the variation of bulk on-shell action can be massaged as
\begin{equation}
\delta S_{bulk}=-\frac{1}{2}\int_{\partial\mathcal {M}}d^3x(\delta
z^\dag_1y_1+\delta z^\dag_2y_2-\delta y^\dag_1z_1-\delta
y^\dag_2z_2-z^\dag_1\delta y_1-z^\dag_2\delta y_2+ y^\dag_1\delta
z_1+y^\dag_2\delta z_2).
\end{equation}
On the other hand, by $\psi=\psi_++\psi_-$, the Lorentz violating
boundary term can be casted into
\begin{equation}
S_{bdy}=-\frac{1}{2}\int_{\partial\mathcal
{M}}d^3x(z^\dag_2y_1+y^\dag_2z_1+z^\dag_1y_2+y^\dag_1z_2).
\end{equation}
Whence the variation of full on-shell action is given by
\begin{eqnarray}\label{non}
\delta S_{bulk}+\delta S_{bdy}&=&-\frac{1}{2}\int_{\partial\mathcal
{M}}d^3x
[\delta(z^\dag_1+z^\dag_2)(y_1+y_2)+\delta(y^\dag_1-y^\dag_2)(z_2-z_1)\nonumber\\
&&+(z^\dag_2-z^\dag_1)\delta(y_1-y_2)+(y^\dag_1+y^\dag_2)\delta(z_1+z_2)]\nonumber\\
&=& -\int_{\partial\mathcal {M}}d^3x (\delta Z^\dag_1
Y_1+Z^\dag_2\delta Y_2+Y^\dag_1\delta Z_1+\delta Y^\dag_2 Z_2),
\end{eqnarray}
where we have defined
$(Y_1,Y_2)=\frac{1}{\sqrt{2}}(y_1+y_2,y_1-y_2)$, and
$(Z_1,Z_2)=\frac{1}{\sqrt{2}}(z_1+z_2,z_2-z_1)$. Hence the corresponding
fermionic source and dual operator are somehow given by $(Z_1,Y_2)$
and $(Y_1,Z_2)$ respectively, and the dimension of operator reads
\begin{equation}
\Delta[Y_1]=\frac{3}{2}+m,\Delta[Z_2]=\frac{3}{2}-m.
\end{equation}

Regarding other possible choices of boundary conditions and how they
are related to one another through the Wilsonian RG flow by the
double trace deformation, please refer to \cite{LT1,LT2}. In the
subsequent section, we will focus only on the above non-relativistic
fixed point, to extract the boundary fermionic correlator from the
bulk dynamics and see how the magnetic dipole coupling affects the
fermionic correlator by numerical calculation.

\section{Holographic non-relativistic fermion with bulk dipole
coupling}\label{mainresult}
\subsection{Holographic setup}
Generically the relevant information regarding the fermionic system
can be read out of its single particle fermionic correlator, namely
the retarded Green function $G_R$. For example, the spectral
function $A(\omega,k)$, which can be measured experimentally by
Angle Resolved Photoemission Spectroscopy(ARPES), is given by the imaginary
part of Tr$G_{R}$. Now we shall show how such a retarded Green
function can be obtained holographically.

To proceed, we would like to start with the Dirac equation
(\ref{Dirac}) in a more general static background, i.e.,
\begin{equation}
ds^2=-g_{tt}(r)dt^2+g_{rr}(r)dr^2+g_{xx}(r)[(dx^1)^2+(dx^2)^2],A_a=A_t(r)(dt)_a.
\end{equation}
 Then we choose the orthogonal normal vector bases as follows
\begin{equation}
(e_t)^a=\frac{1}{\sqrt{g_{tt}}}(\frac{\partial}{\partial
t})^a,(e_i)^a=\frac{1}{\sqrt{g_{xx}}}(\frac{\partial}{\partial
x^i})^a,(e_r)^a=\frac{1}{\sqrt{g_{rr}}}(\frac{\partial}{\partial
r})^a,
\end{equation}
from which the non-vanishing components of spin connection can be
obtained as
\begin{equation}
(\omega_{tr})_a=-(\omega_{rt})_a=-\frac{\partial_r\sqrt{g_{tt}}}{\sqrt{g_{rr}}}(dt)_a,(\omega_{ir})_a=-(\omega_{ri})_a=\frac{\partial_r\sqrt{g_{xx}}}{\sqrt{g_{rr}}}(dx^i)_a.
\end{equation}
Next let $\psi=(-h)^{-\frac{1}{4}}\varphi$, then the bulk Dirac
equation can be expressed as
\begin{equation}
\frac{\Gamma^r\partial_r\varphi}{\sqrt{g_{rr}}}+\frac{\Gamma^t(\partial_t-iqA_t)\varphi}{\sqrt{g_{tt}}}+\frac{\Gamma^i\partial_i\varphi}{\sqrt{g_{xx}}}
-m\varphi-\frac{ip\Gamma^{rt}\partial_rA_t}{2\sqrt{g_{tt}g_{rr}}}\varphi=0.
\end{equation}
By the rotation symmetry in the spatial directions, without loss of
generality, we shall let $\varphi=e^{-i\omega
t+ikx^1}\tilde{\varphi}$, thus the Dirac equation reduces to
\begin{equation}
\frac{\sqrt{g_{xx}}}{\sqrt{g_{rr}}}(\Gamma^r\partial_r-m\sqrt{g_{rr}}-\frac{ip\Gamma^{rt}\partial_rA_t}{2\sqrt{g_{tt}}})\tilde{\varphi}+(-iu\Gamma^t+ik\Gamma^1)\tilde{\varphi}=0,
\end{equation}
where
\begin{equation}
u=\frac{\sqrt{g_{xx}}}{\sqrt{g_{tt}}}(\omega+qA_t).
\end{equation}
Set $\tilde{\varphi}=\left(
                       \begin{array}{c}
                         \tilde{\varphi}_1 \\
                         \tilde{\varphi}_2\\
                       \end{array}
                     \right)$,
                     then with our representation of Gamma
                     matrices, the equation of motion can be further simplified as
\begin{equation}
\frac{\sqrt{g_{xx}}}{\sqrt{g_{rr}}}(\partial_r+m\sqrt{g_{rr}}\sigma_3)\tilde{\varphi}_I=\Big[i\sigma_2u+[(-1)^I
k-\frac{p\sqrt{g_{xx}}\partial_rA_t}{\sqrt{g_{tt}g_{rr}}}]\sigma_1\Big]\tilde{\varphi}_I,
\end{equation}
with $I=1,2$. Moreover, by $\tilde{\varphi}_I=\left(
                     \begin{array}{c}
                       \tilde{y}_I \\
                       \tilde{z}_I \\
                     \end{array}
                   \right)$, the above equation of motion gives rise to the following flow equation, i.e.,

                   \begin{equation}
                   \frac{\sqrt{g_{xx}}}{\sqrt{g_{rr}}}\partial_r\xi_I=-2m\sqrt{g_{xx}}\xi_I+[u-\frac{p\sqrt{g_{xx}}\partial_rA_t}{\sqrt{g_{tt}g_{rr}}}+(-1)^Ik]+[u+\frac{p\sqrt{g_{xx}}\partial_rA_t}{\sqrt{g_{tt}g_{rr}}}-(-1)^Ik]\xi_I^2
                   \end{equation}
where $\xi_I=\frac{\tilde{y}_I}{\tilde{z}_I}$. Now plug the
particular background (\ref{blackhole}) into the above equations, we
end up with
\begin{equation}\label{eom}
(r^2\sqrt{f}\partial_r+rm\sigma_3)\tilde{\varphi}_I=\Big[\frac{i\sigma_2}{\sqrt{f}}[\omega+qg_FQ(1-\frac{1}{r})]+[(-1)^Ik-\frac{pg_FQ}{r}]\Big]\tilde{\varphi}_I
\end{equation}
for the equation of motion, and
\begin{equation}\label{flow}
r^2\sqrt{f}\partial_r\xi_I=-2mr\xi_I+[v_-+(-1)^Ik]+[v_+-(-1)^Ik]\xi_I^2
\end{equation}
for the flow equation, where
\begin{equation}
v_\pm=\frac{1}{\sqrt{f}}[\omega+qg_FQ(1-\frac{1}{r})]\pm\frac{pg_FQ}{r}.
\end{equation}
 Whence near the boundary, namely when $r$ goes to the infinity,
$\tilde{\varphi}_I$ behave in the following way, i.e.,
\begin{equation}
\tilde{\varphi}_I\rightarrow c_I r^m\left(
                                                  \begin{array}{c}
                                                    0 \\
                                                    1 \\
                                                  \end{array}
                                                \right)+d_I
                                                r^{-m}\left(
                                                        \begin{array}{c}
                                                          1 \\
                                                          0 \\
                                                        \end{array}
                                                      \right).
\end{equation}
The ratio $G_I=\frac{d_I}{c_I}$ can be fixed by imposing the
in-falling boundary condition for $\tilde{\varphi}$ at the horizon,
where $\tilde{\varphi}_I$ behave as
\begin{equation}
\tilde{\varphi}_I\propto \left(
                                                  \begin{array}{c}
                                                    i \\
                                                    1 \\
                                                  \end{array}
                                                \right)e^{-i\omega
                                                r_*}
 \end{equation}
 with $r_*=\int\frac{dr}{r^2f}$. Alternatively, this ratio can also be obtained in a more convenient way as
                               \begin{equation}
                               G_I=\lim_{r\rightarrow\infty}r^{2m}\xi_I,
                               \end{equation}
 by solving the flow equation (\ref{flow}) with the boundary condition at the horizon
 \begin{equation}
 \xi_I=i.
 \end{equation}
 Then by the recipe of AdS/CFT, it follows from the variation of
 on-shell action (\ref{non}) that at the non-relativistic fixed point the retarded fermionic Green
correlator can be extracted from the following relation, i.e.,
\begin{eqnarray}
\left(
               \begin{array}{c}
                 D_1 \\
                 C_2 \\
               \end{array}
             \right)
=G_R\left(
               \begin{array}{c}
                 C_1 \\
                 D_2 \\
               \end{array}
             \right)=\left(
                                 \begin{array}{cc}
                                     \alpha & \beta \\
                                     \gamma & \eta \\
                                   \end{array}
                                 \right)
\left(
               \begin{array}{c}
                 C_1 \\
                 D_2 \\
               \end{array}
             \right)
\end{eqnarray}
with $(D_1,D_2)=\frac{1}{\sqrt{2}}(d_1+d_2,d_1-d_2)$ and
$(C_1,C_2)=\frac{1}{\sqrt{2}}(c_1+c_2,c_2-c_1)$. To be more precise,
from such a relation, we have
\begin{eqnarray}
\frac{\alpha}{G_1}+\beta=1,\ \ \ \frac{\alpha}{G_2}-\beta=1,\nonumber\\
\gamma+\eta G_1=-1,\ \ \ \ \gamma-\eta G_2=1,
\end{eqnarray}
which gives $G_R$ in terms of $G_I$ as\footnote{It is noteworthy that $G_I$ are simply the components of  diagonal retarded Green function in the standard quantization. An intriguing question is how to understand (\ref{rg}) directly from the RG flow induced by the double trace deformation. }
\begin{eqnarray}\label{rg}
G_R=\left(
               \begin{array}{cc}
                 \frac{2G_1G_2}{G_1+G_2} & \frac{G_1-G_2}{G_1+G_2} \\
                 \frac{G_1-G_2}{G_1+G_2} & \frac{-2}{G_1+G_2} \\
               \end{array}
             \right).
\end{eqnarray}
 Consequently, $det(G_R)=-1$,  and the trace along with the eigenvalues of $G_R$  can be
 worked out as
\begin{eqnarray}\label{e2}
&&\lambda_\pm=\frac{G_1G_2-1\pm\sqrt{1+G_1^2+G_2^2+G_1^2G_2^2}}{G_1+G_2}\\
&&\mathtt{Tr}G_R=\lambda_++\lambda_-=\frac{2G_1G_2-2}{G_1+G_2}.
\end{eqnarray}
 It follows from
the flow equation (\ref{flow}) that $G_1$ and $G_2$ are related to
each other as
\begin{equation}
G_2(\omega,k)=G_1(\omega,-k).
\end{equation}
Therefore both of the trace and eigenvalues of our retarded Green
function are invariant under the transformation $k\rightarrow -k$ as
it should be, guaranteed by the rotation symmetry mentioned above.
In addition, by the flow equation, we also have
\begin{equation}
G_1(\omega,k;g_F)=-G_2^*(-\omega,k;-g_F),
\end{equation}
or equivalently
\begin{equation}
G_1(\omega,k;q,p)=-G_2^*(-\omega,k;-q,-p).
\end{equation}
To make our life easier, in what follows, we will set $g_F=1$ and
work exclusively with the case of $m=0$, where the flow equation
further implies
\begin{equation}\label{easy}
G_1(\omega,k;p)=-\frac{1}{G_2(\omega,k;-p)}.
\end{equation}
So it is essentially enough to restrict ourselves to non-negative
$k$, $q$ and $p$.
\subsection{Numerical results}
In the following numerical calculations, we shall focus exclusively on the probe fermion in the zero temperature soup, which can be achieved by setting $Q=\sqrt{3}$. Then the chemical potential is given by $\mu=\sqrt{3}q$.
\begin{figure}
\centering
\includegraphics[width=.48\textwidth]{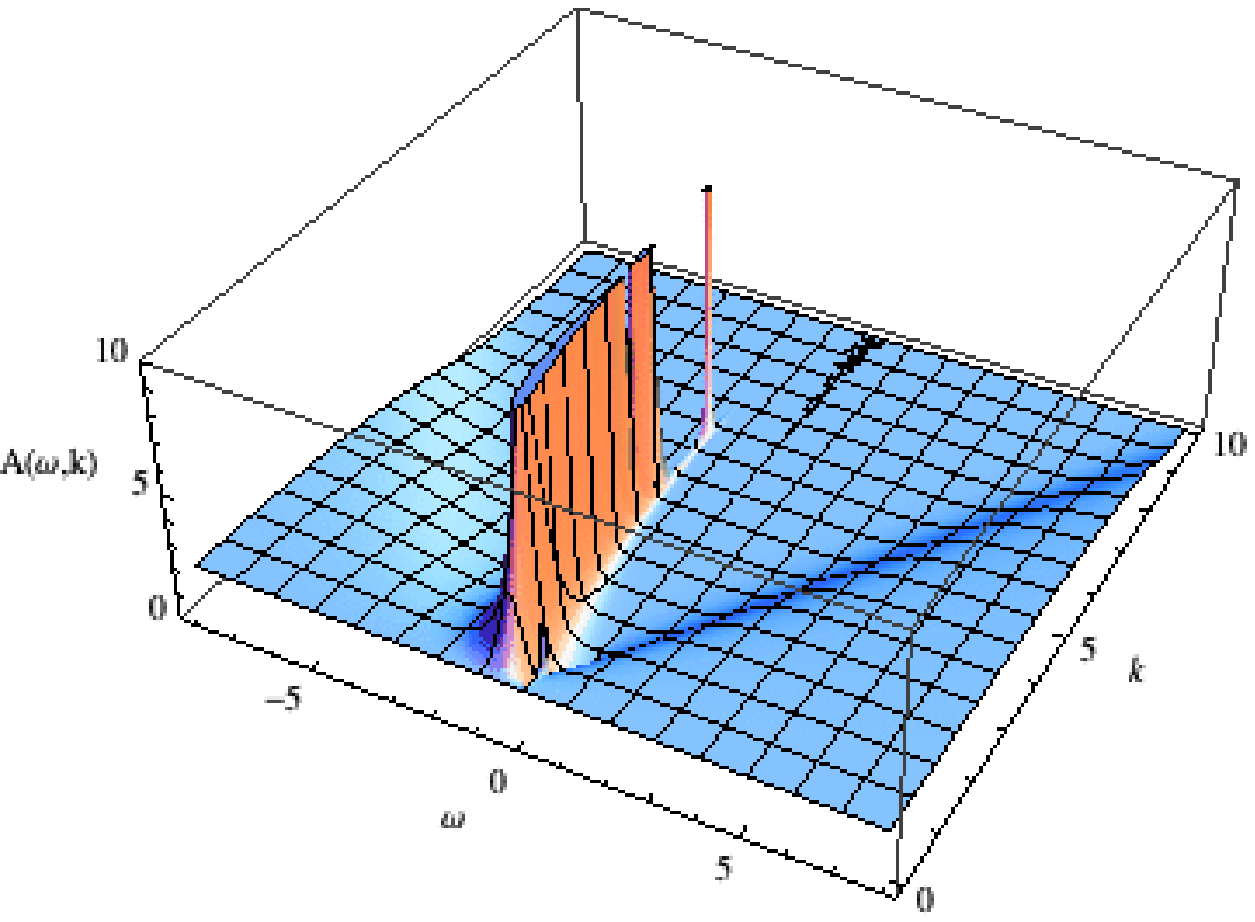}
      \includegraphics[width=.48\textwidth]{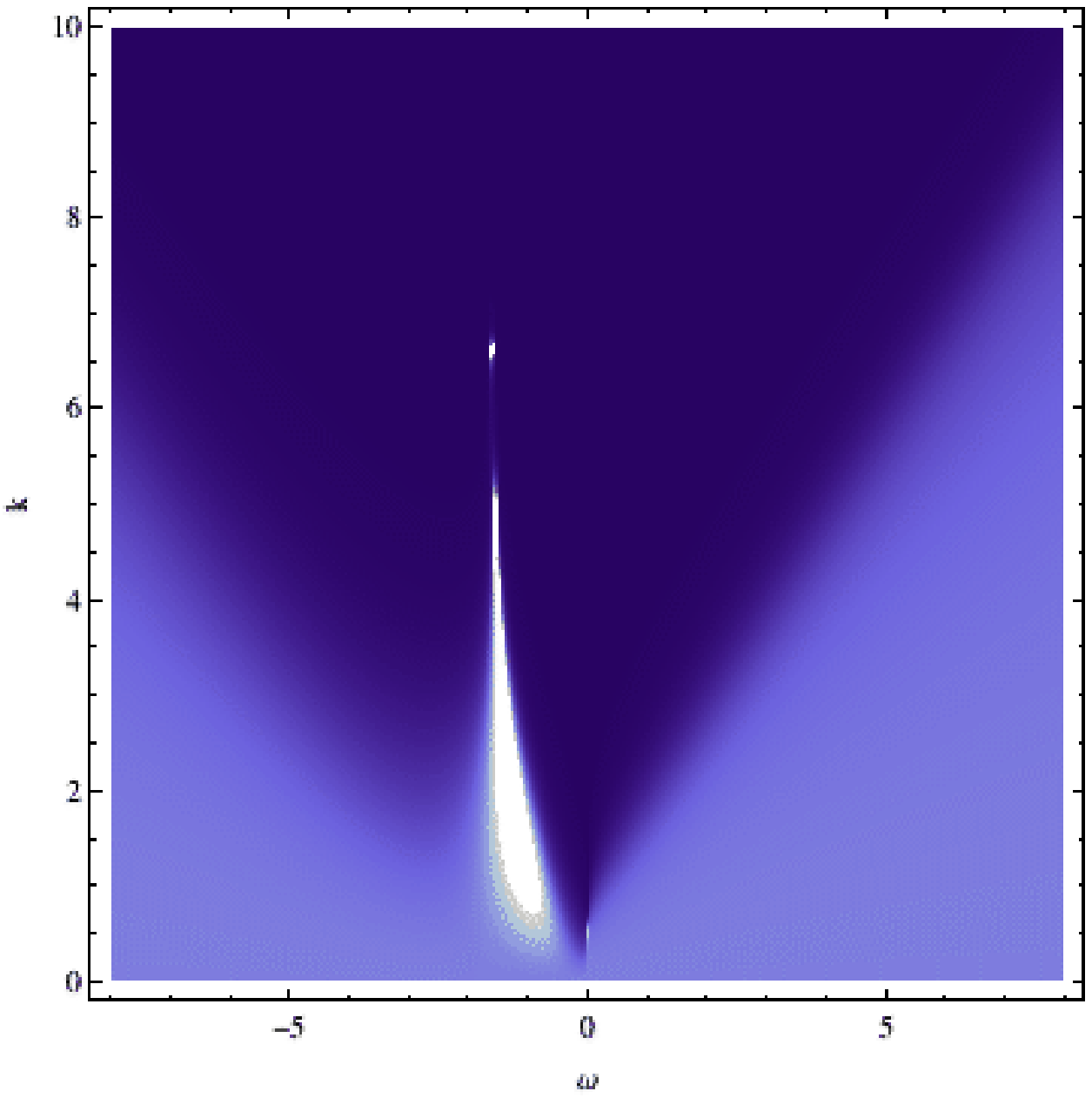}
       \caption{The 3d and density plots of spectral function  for the case of $q=1$ and $p=0$. }
       \label{p0q1}
\end{figure}

\begin{figure}
\centering
\includegraphics[width=.48\textwidth]{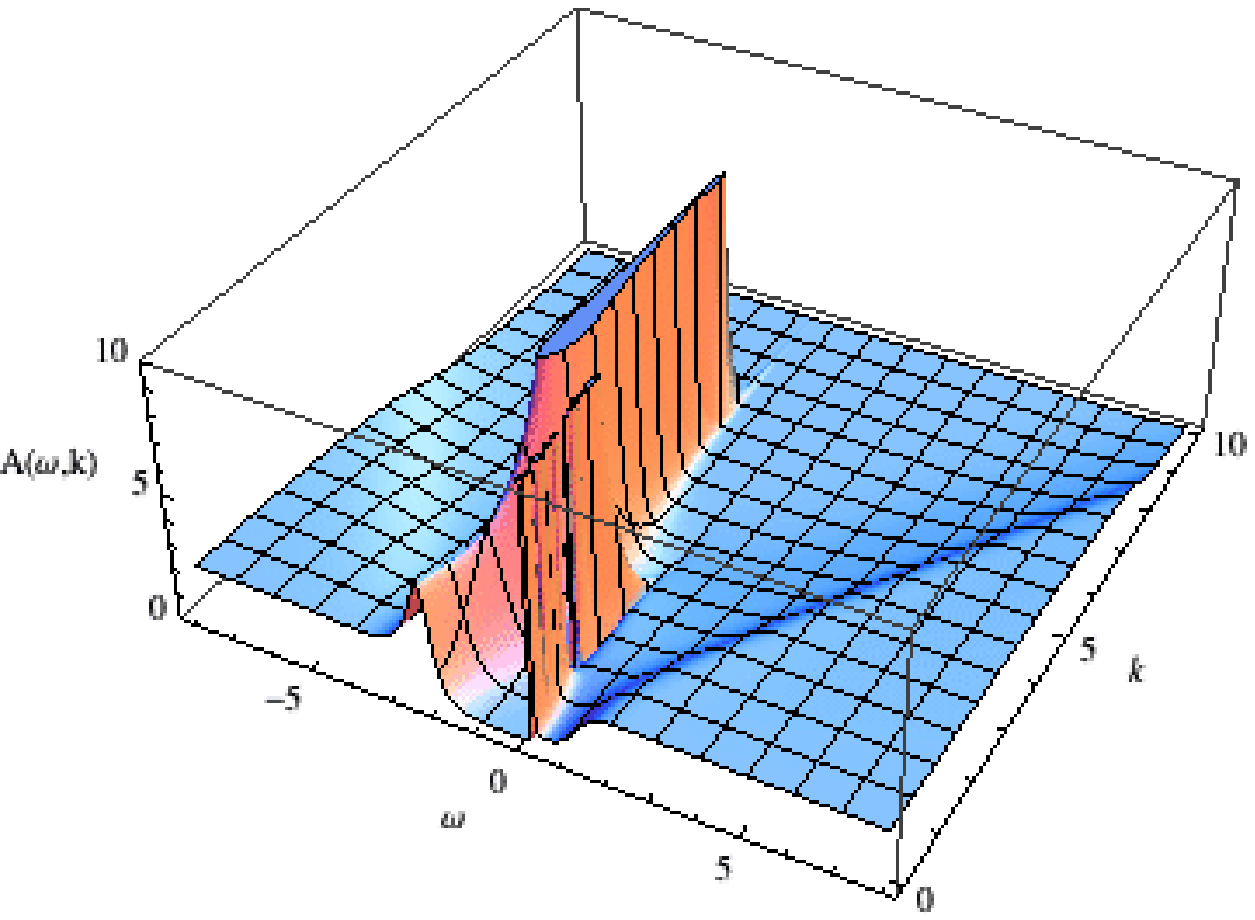}
      \includegraphics[width=.48\textwidth]{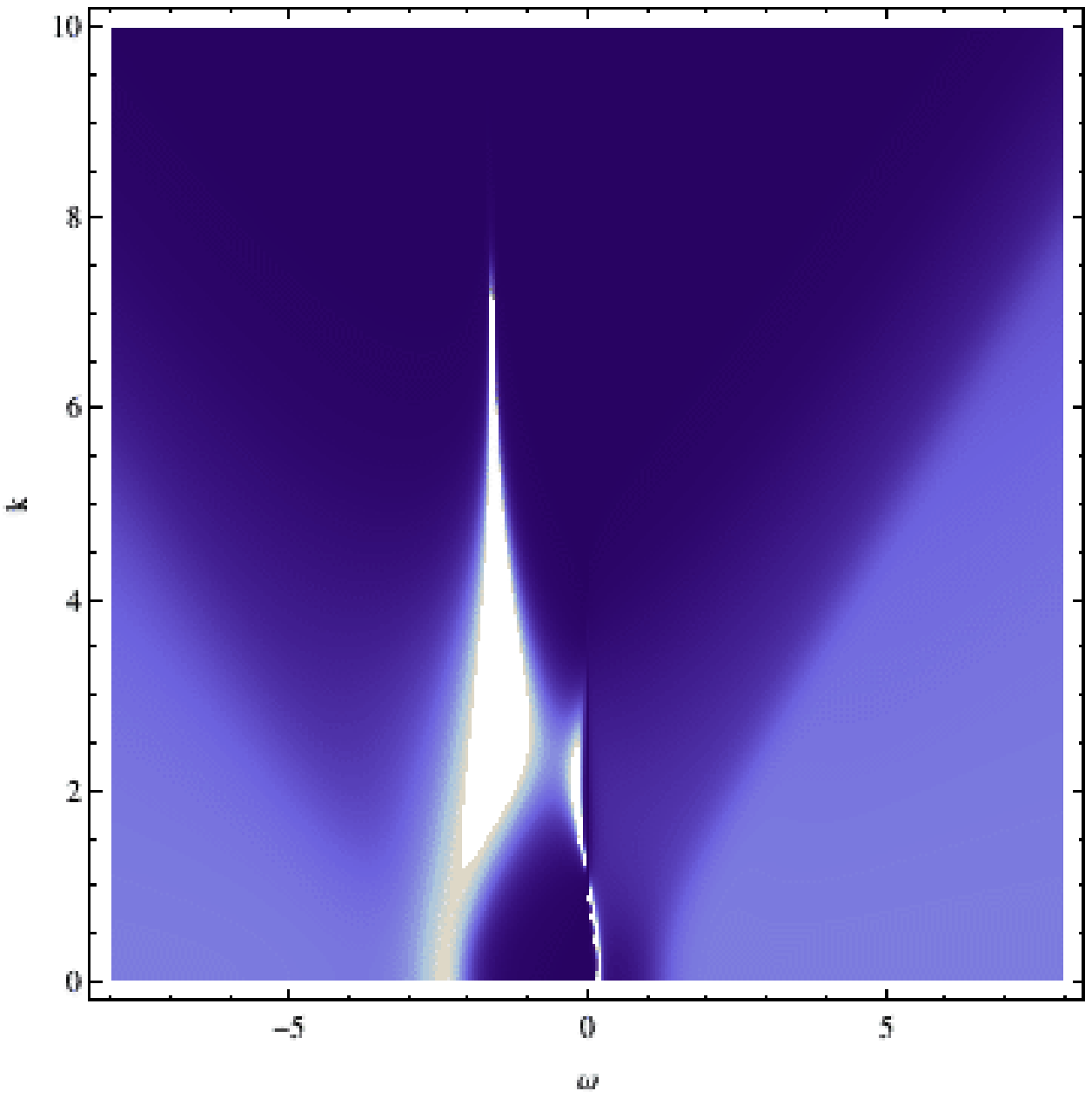}
       \caption{The 3d and density plots of spectral function  for the case of $q=1$ and $p=2$.     }
       \label{p2q1}
\end{figure}

\begin{figure}
\centering
\includegraphics[width=.48\textwidth]{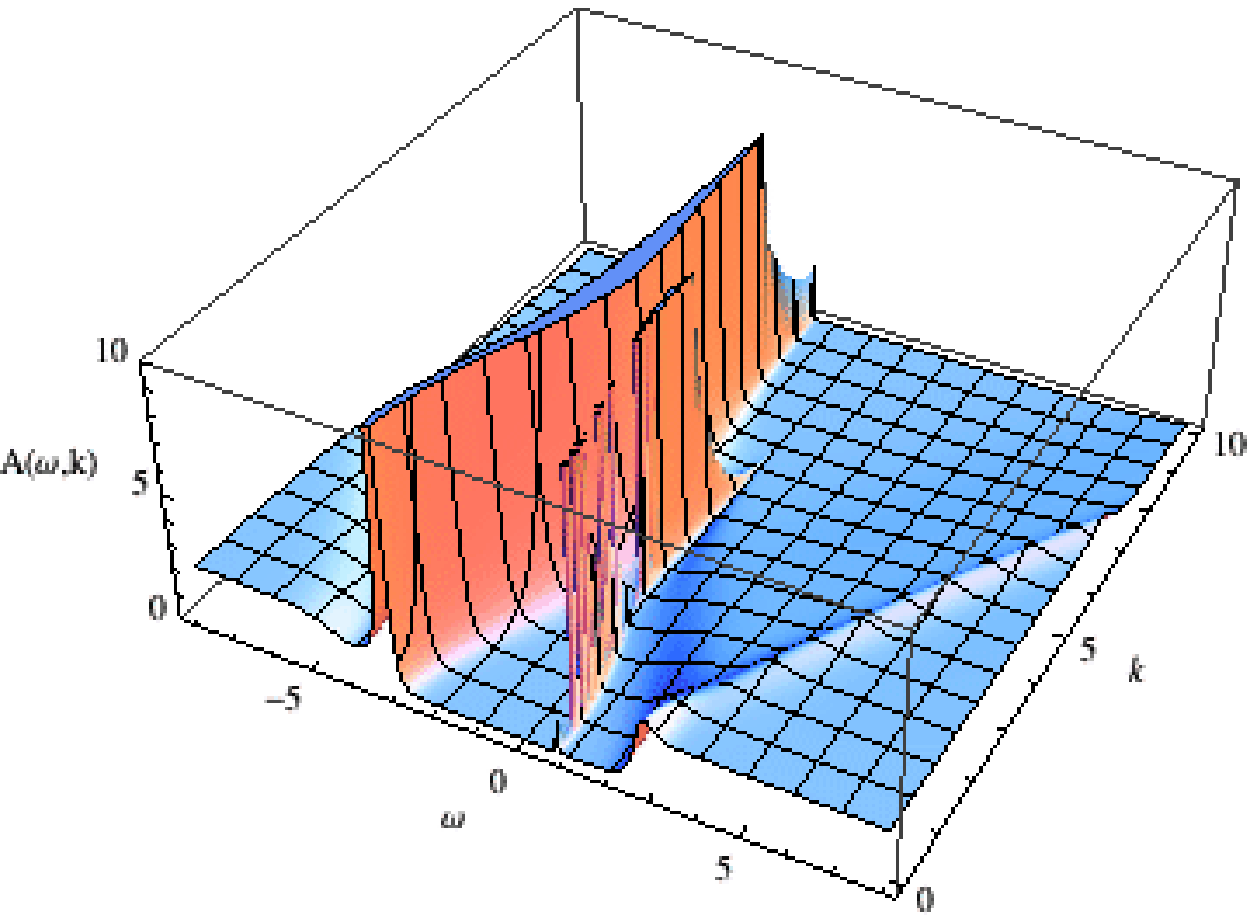}
      \includegraphics[width=.48\textwidth]{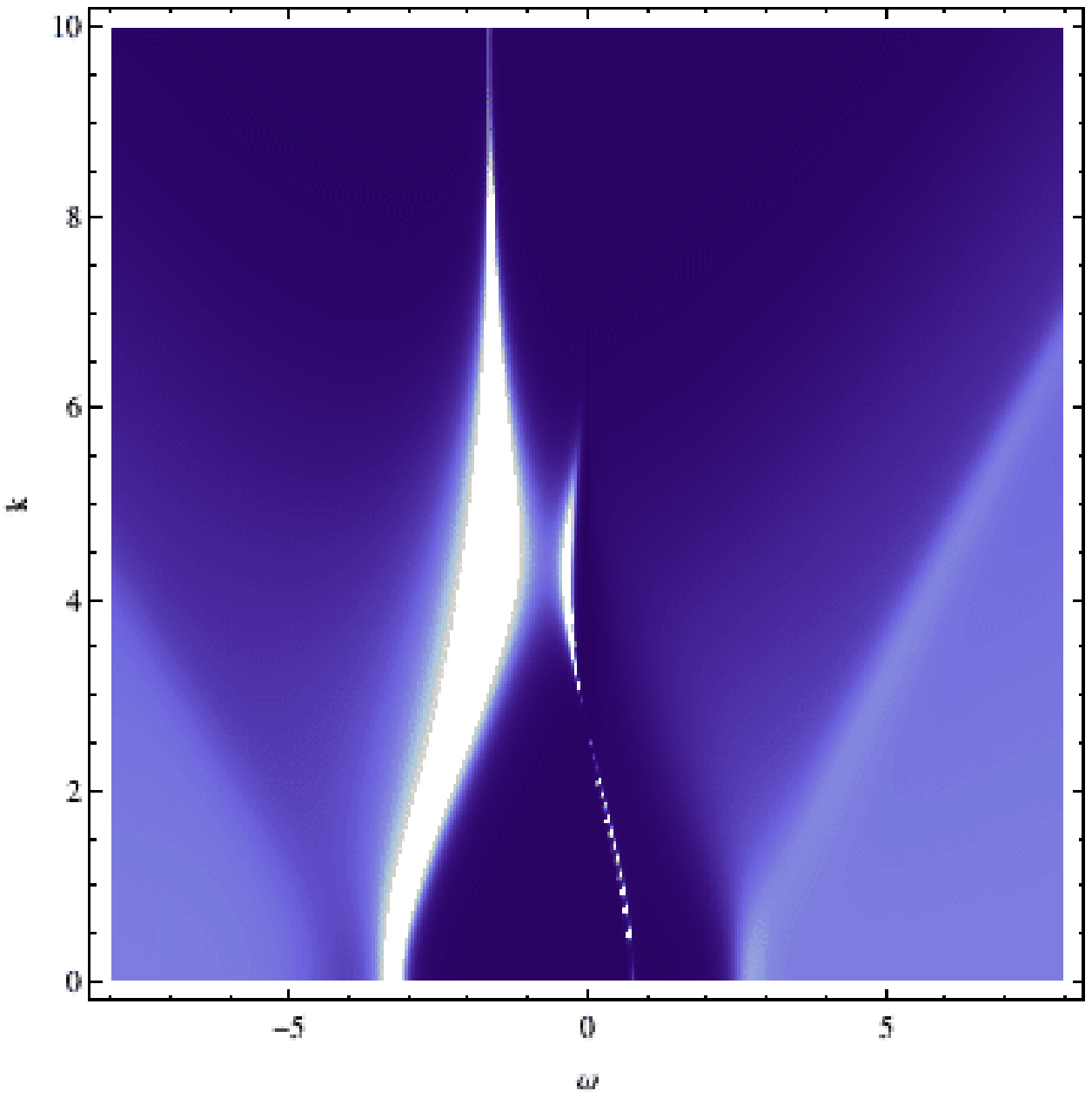}
       \caption{ The 3d and density plots of spectral function  for the case of $q=1$ and $p=4$.   }
       \label{p4q1}
\end{figure}
\begin{figure}
\centering
\includegraphics[width=.48\textwidth]{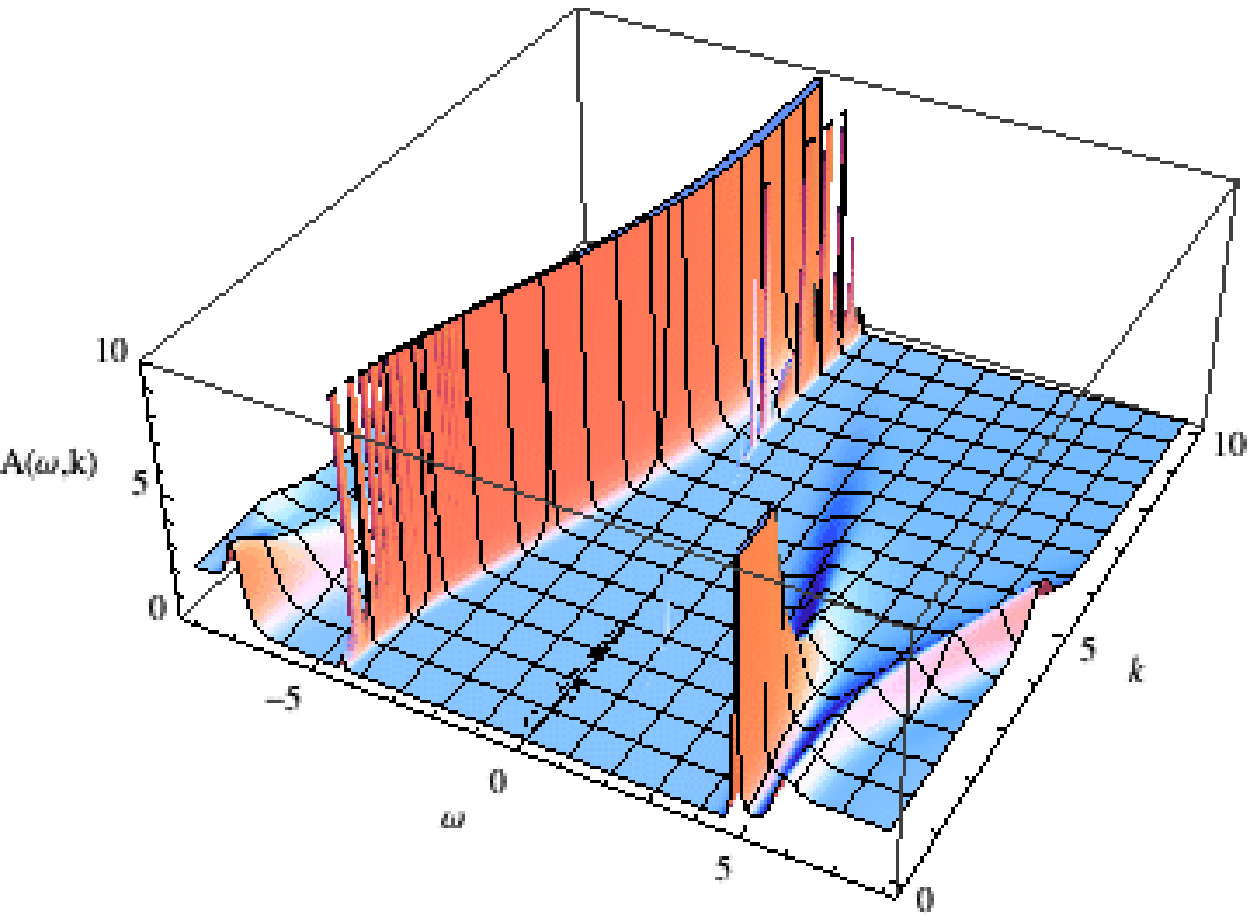}
      \includegraphics[width=.48\textwidth]{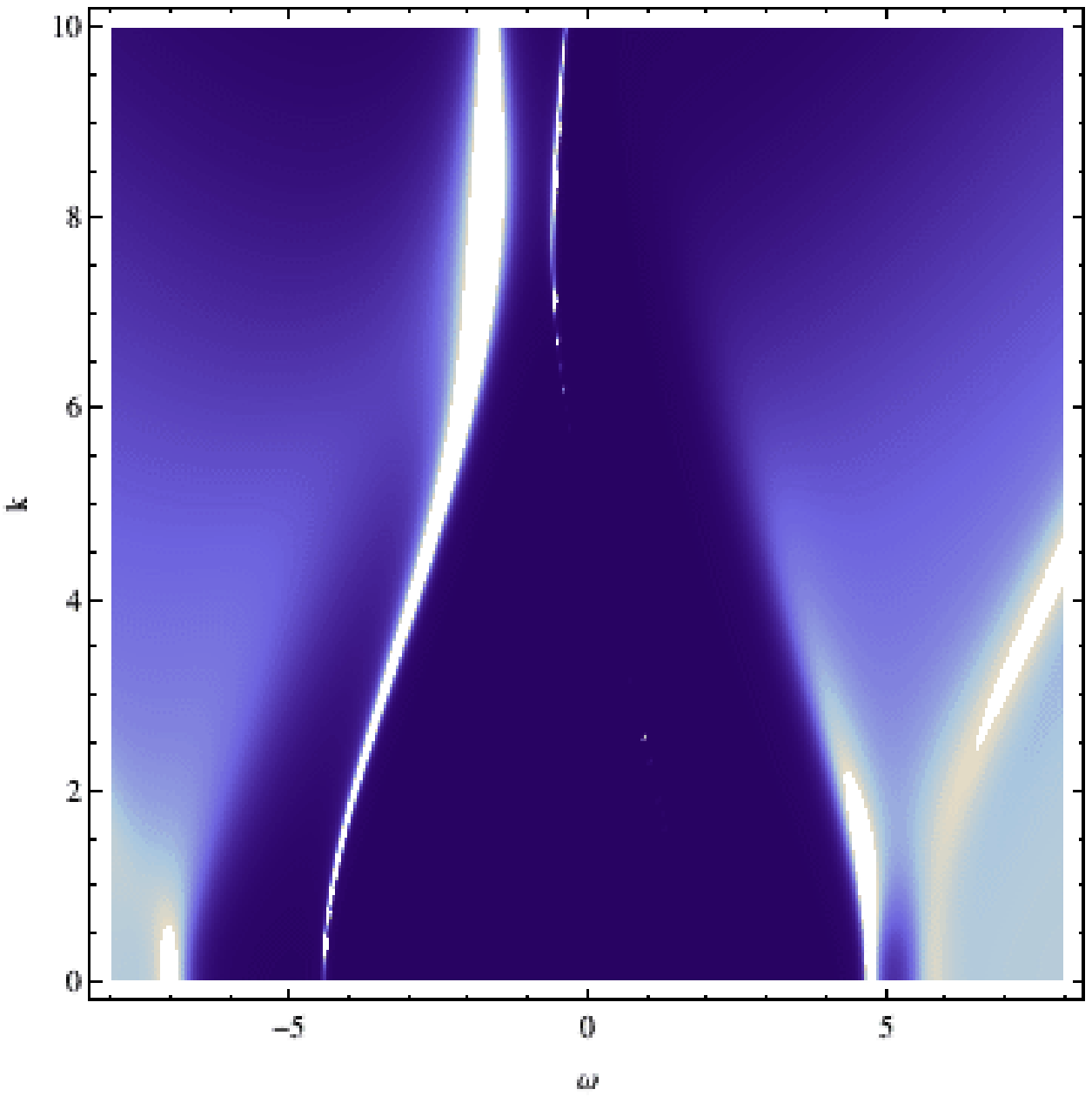}
       \caption{ The 3d and density plots of spectral function  for the case of $q=1$ and $p=8$.   }
       \label{p8q1}
\end{figure}
\begin{figure}
\includegraphics[width=.96\textwidth]{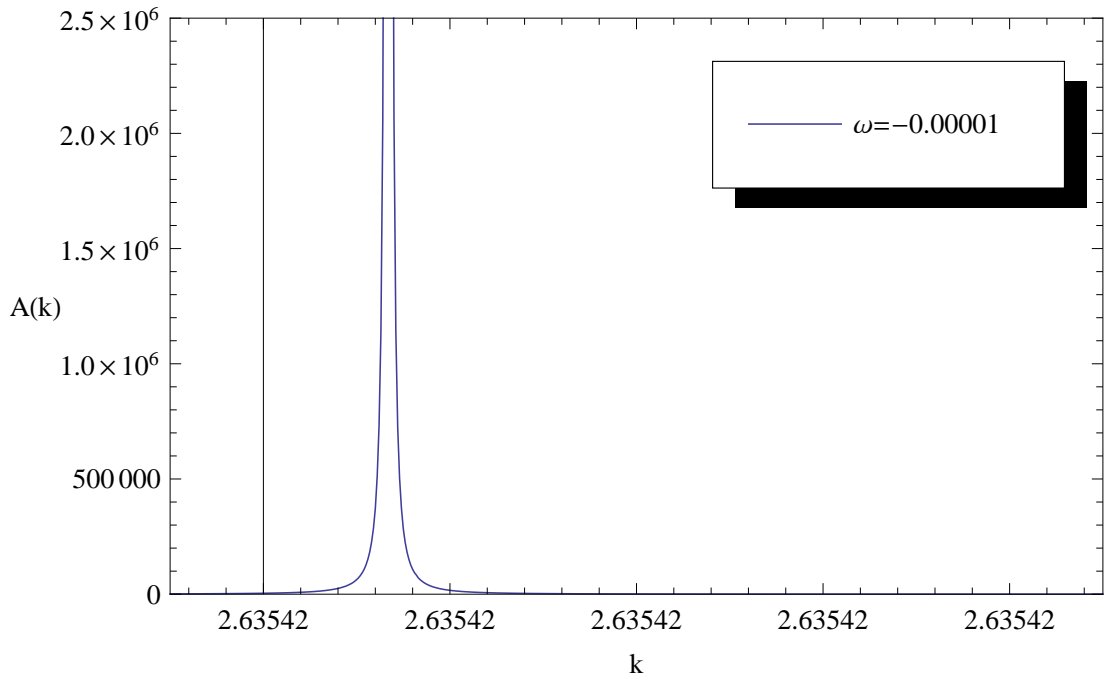}
 \caption{The Fermi momentum can be fixed to the 5th digit as $k_F=2.63542$ by taking $\omega=-0.00001$ for the case of  $q=1$ and $p=4$.   }
       \label{fs}
\end{figure}

\begin{table}[htbp]
   \centering
   \begin{tabular}{|c|c|c|c|c|}
   \hline
     q=1&p=0&p=2&p=4&p=8\\
     \hline
     $k_F$&No&1.08188&2.63542&4.87894\\
     \hline
   \end{tabular}
   \caption{The Fermi momentum for the case of $q=1$.}
   \label{fm}
\end{table}
We start by fixing $q=1$ but varying $p$. The corresponding
numerical results are plotted in Figures \ref{p0q1}, \ref{p2q1} ,
\ref{p4q1},  and \ref{p8q1}, where the 3d and density plots are
drawn for the spectral function on the left and right respectively.
First, albeit with the different choice of Gamma matrices the result
presented in Figure \ref{p0q1} is consistent with that obtained in
\cite{LT1}, where instead of the spectral function, Im$\lambda_+$
and Im$\lambda_-$ are plotted individually\footnote{As pointed out
to us by Robert Leigh,  Im$\lambda_+$ and Im$\lambda_-$ can also be
measured by the so called spin polarized ARPES. The reason why we
work only with the spectral function lies in the fact that unlike
the trace of retarded Green function the expression for two
eigenvalues generically involves the square root, as shown in
(\ref{e2}). Such a square root makes it less easy to separate these
two eigenvalues by our numerical calculations. The only exception is
the case of $m=0$ and $p=0$, where the two eigenvalues can be
explicitly massaged as $\lambda_+=\frac{G_1-1}{G_1+1}$ and
$\lambda_-=\frac{1+G_1}{1-G_1}$ by using (\ref{easy}). }. This can
be regarded as sort of consistency check on our numerics. Second,
with the finite chemical potential the infinite flat band get mildly
dispersed. In the large momentum limit, the band becomes
asymptotically flat. This is reasonable as at large momenta the
corresponding modes sit outside of the light cone and can not decay.
Furthermore, as one goes to the large momentum limit, the flat band
is always shifted to $\omega=-\sqrt{3}$, independent of the specific
value of $p$, which arises because the frequency is measured with
respect to the chemical potential. But  the larger becomes the value
of $p$, the larger becomes the momentum at which the peak of flat
band tends to be sharp, which is somehow related to something more
interesting occurring in the region of small momenta. The infinite
band, which is destroyed or depleted from zero momentum up to the
momentum of the order of the finite chemical potential by colliding
our probe fermion with the relativistic soup, is gradually recovered
to extend down to zero momentum as one pushes the bulk dipole
coupling up. In particular, at the momenta lower than the value of
$p$ the flat band disperses the other way than it does at large
momenta. This implies that it costs more energy to excite small
momentum modes than intermediate momentum modes, which is also
consistent with the phenomenon that the peak of flat band becomes
sharper and sharper at small momenta as $p$ is increased. Moreover,
as shown in Figure \ref{p8q1}, some other gap states, which are
nearly flat  finite bands,  are generated at small momenta when the
value of $p$ is large enough.

On the other hand, as $p$ increases, the Fermi surface starts to
show up  at some point with the Fermi momentum increasing. To be
more precise,  as demonstrated in Figure \ref{fs}, we can identify
the location of Fermi surface to the 5th digit using the fact that
the location of peak approaches the Fermi surface $k_F$ in the limit
$\omega\rightarrow 0$. The corresponding results are listed in Table
\ref{fm}\footnote{Note that in Figure \ref{p8q1} the apparent
disappearance of the peak around $\omega=0$ is sort of numerical
artifact. It arises because the peak becomes sharper as one
approaches $\omega=0$.}. It is noteworthy that such a situation is
different from what is happening to the relativistic fixed points,
where the Fermi surface disappears and instead the gap forms when
the bulk dipole coupling is large enough\cite{ELP}.

\begin{figure}
\centering
\includegraphics[width=.48\textwidth]{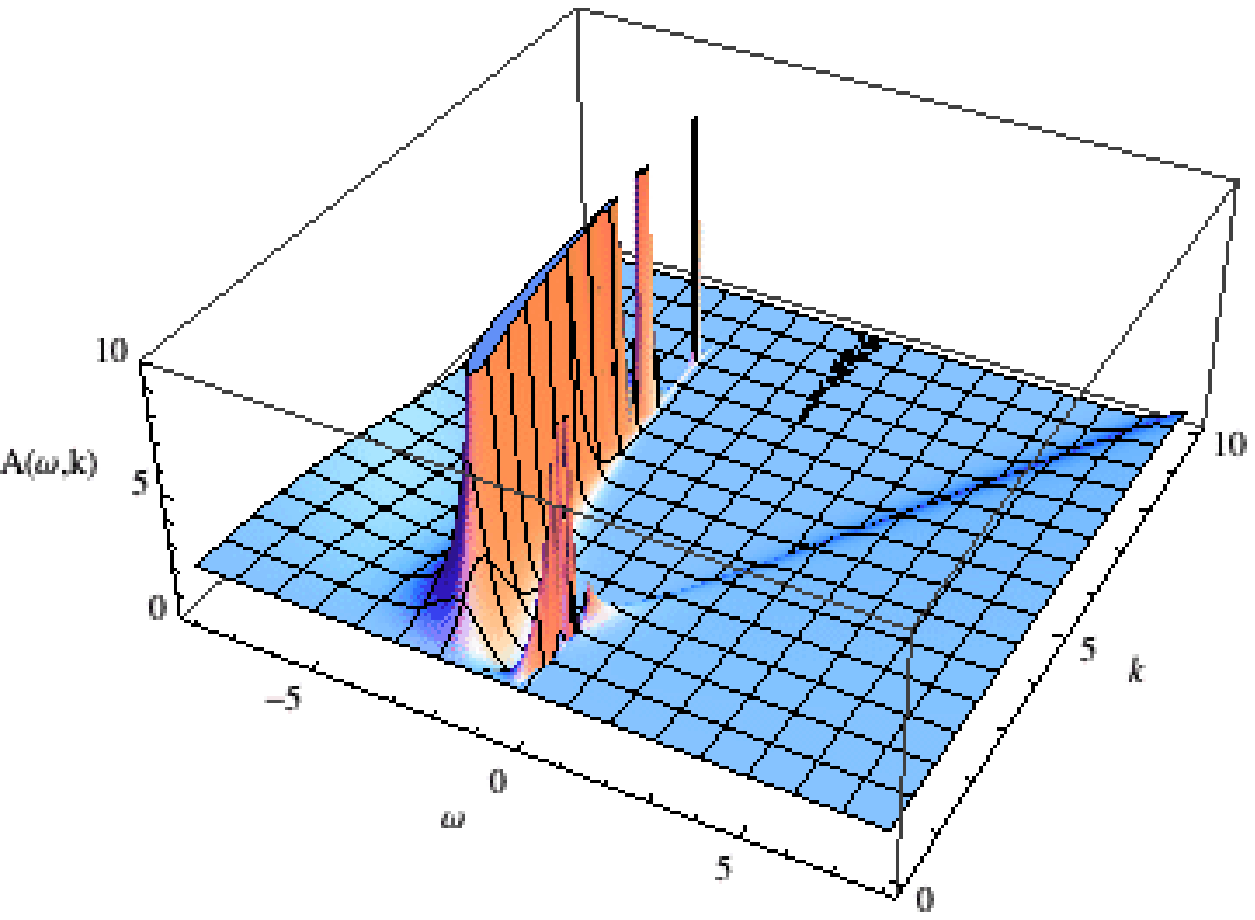}
      \includegraphics[width=.48\textwidth]{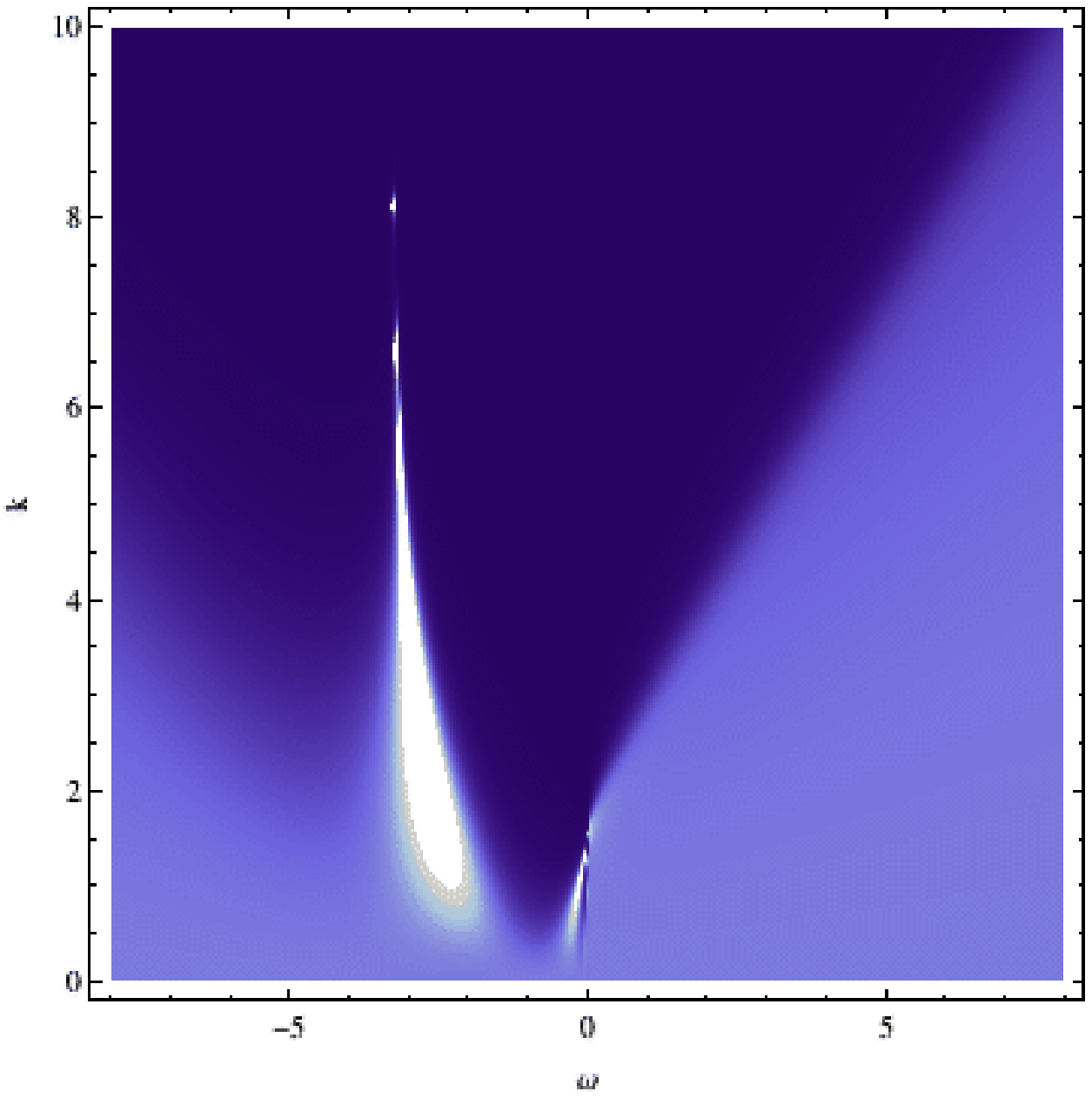}
       \caption{The 3d and density plots of spectral function for the case of $q=2$ and $p=0$, where the Fermi surface shows up with the Fermi momentum $k_F=1.53521$.     }
       \label{p0q2}
\end{figure}
\begin{figure}
\centering
\includegraphics[width=.48\textwidth]{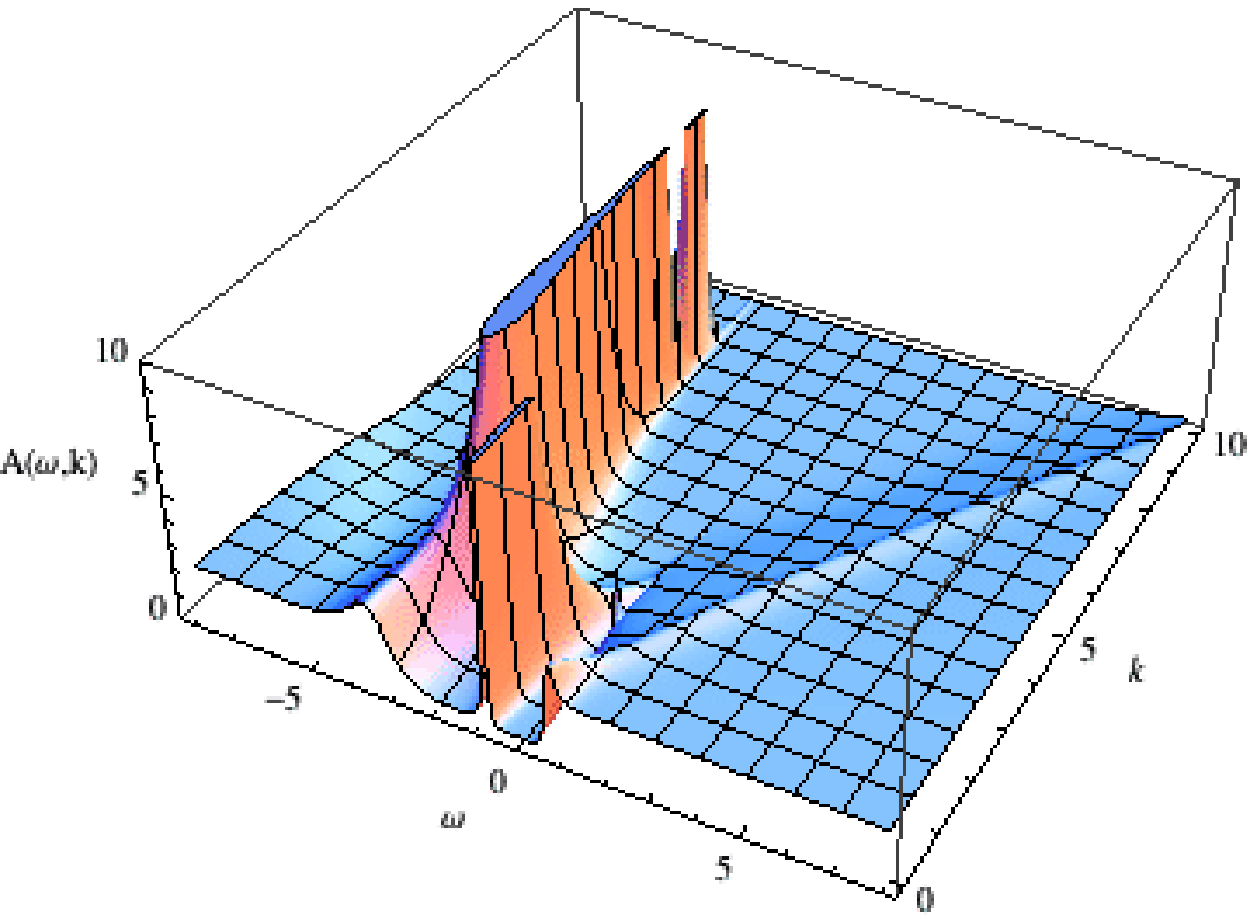}
      \includegraphics[width=.48\textwidth]{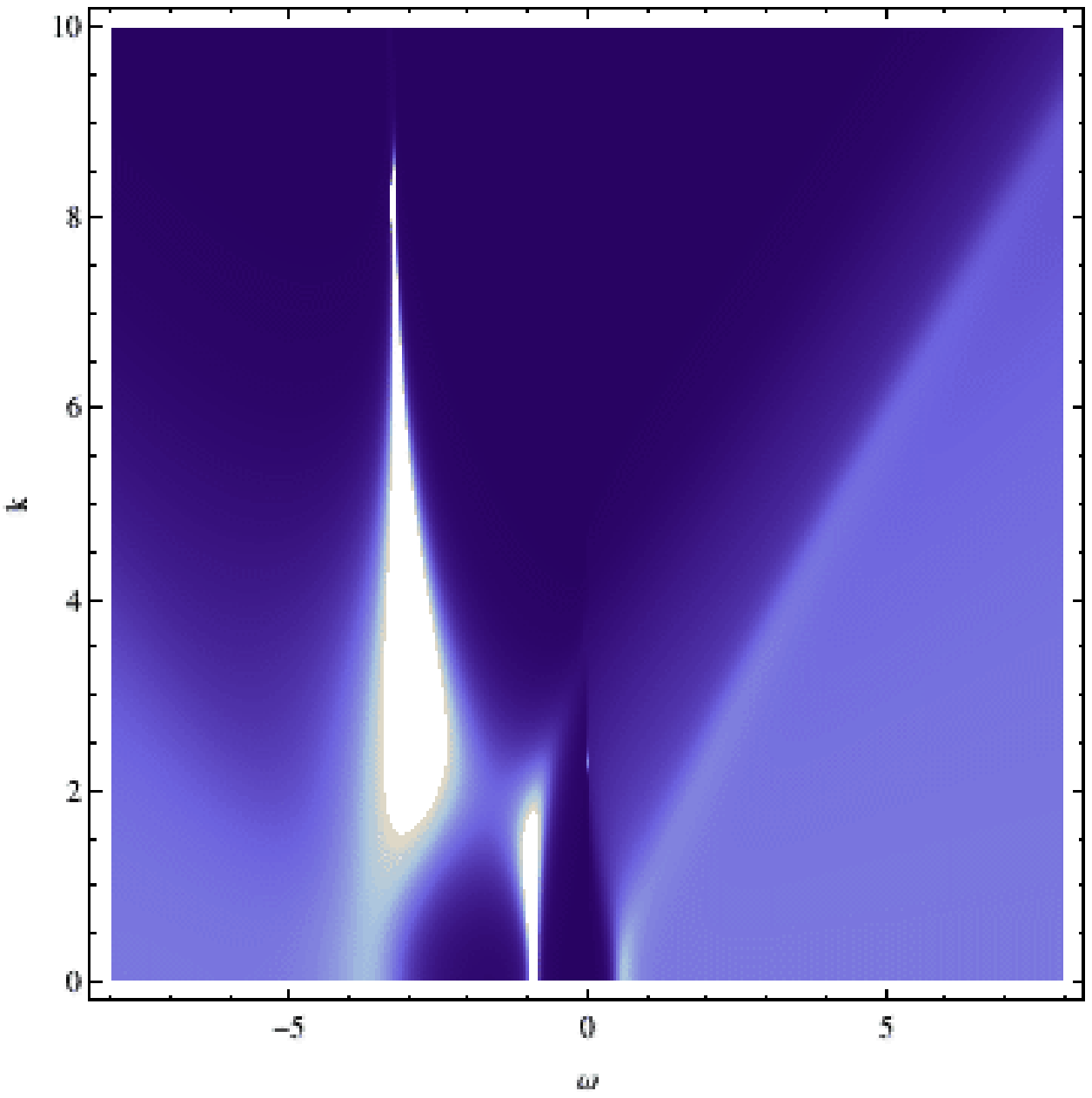}
       \caption{The 3d and density plots of spectral function for the case of $q=2$ and $p=2$, where the Fermi surface disappears and a flat gap sits at $\omega=-0.85$.     }
       \label{p2q2}
\end{figure}
\begin{figure}
\centering
\includegraphics[width=.48\textwidth]{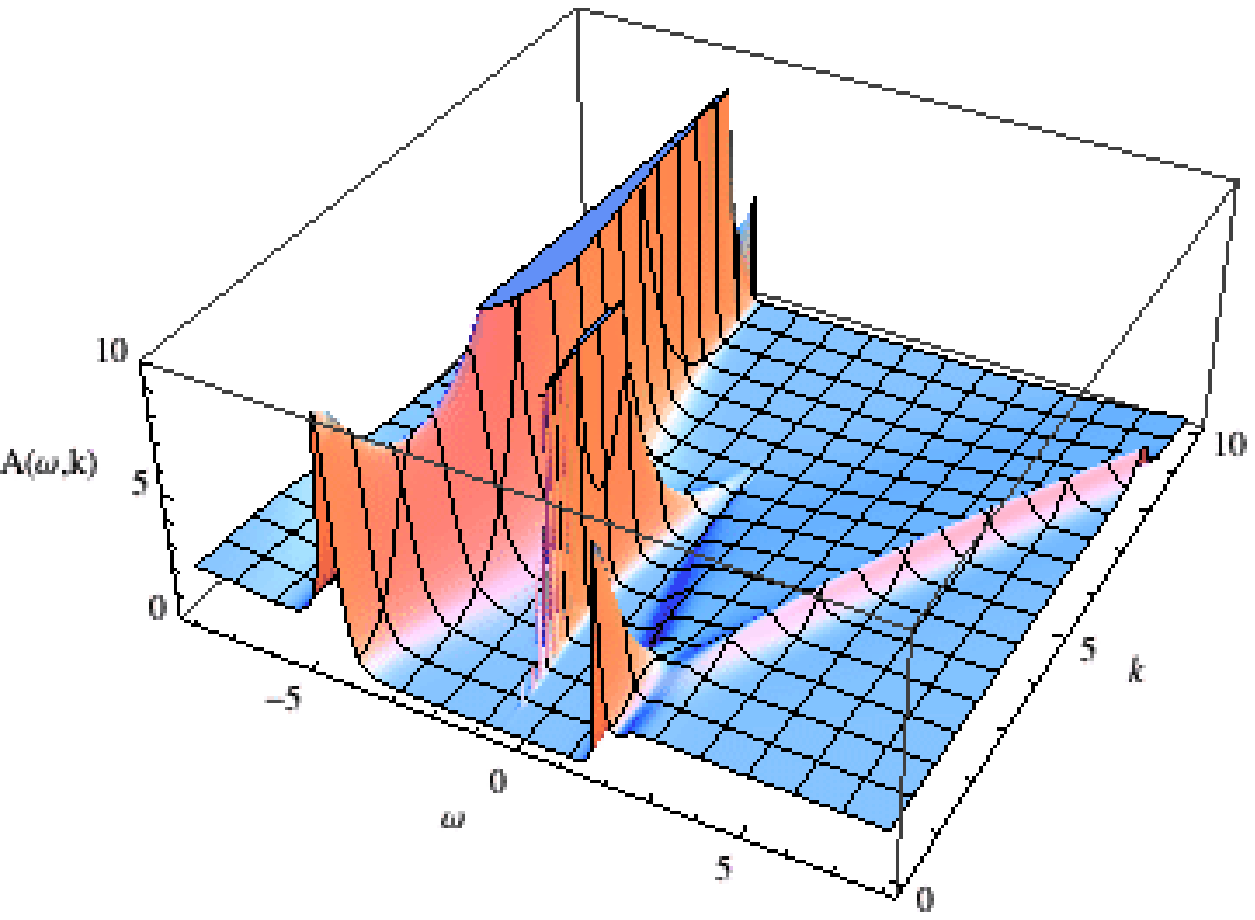}
      \includegraphics[width=.48\textwidth]{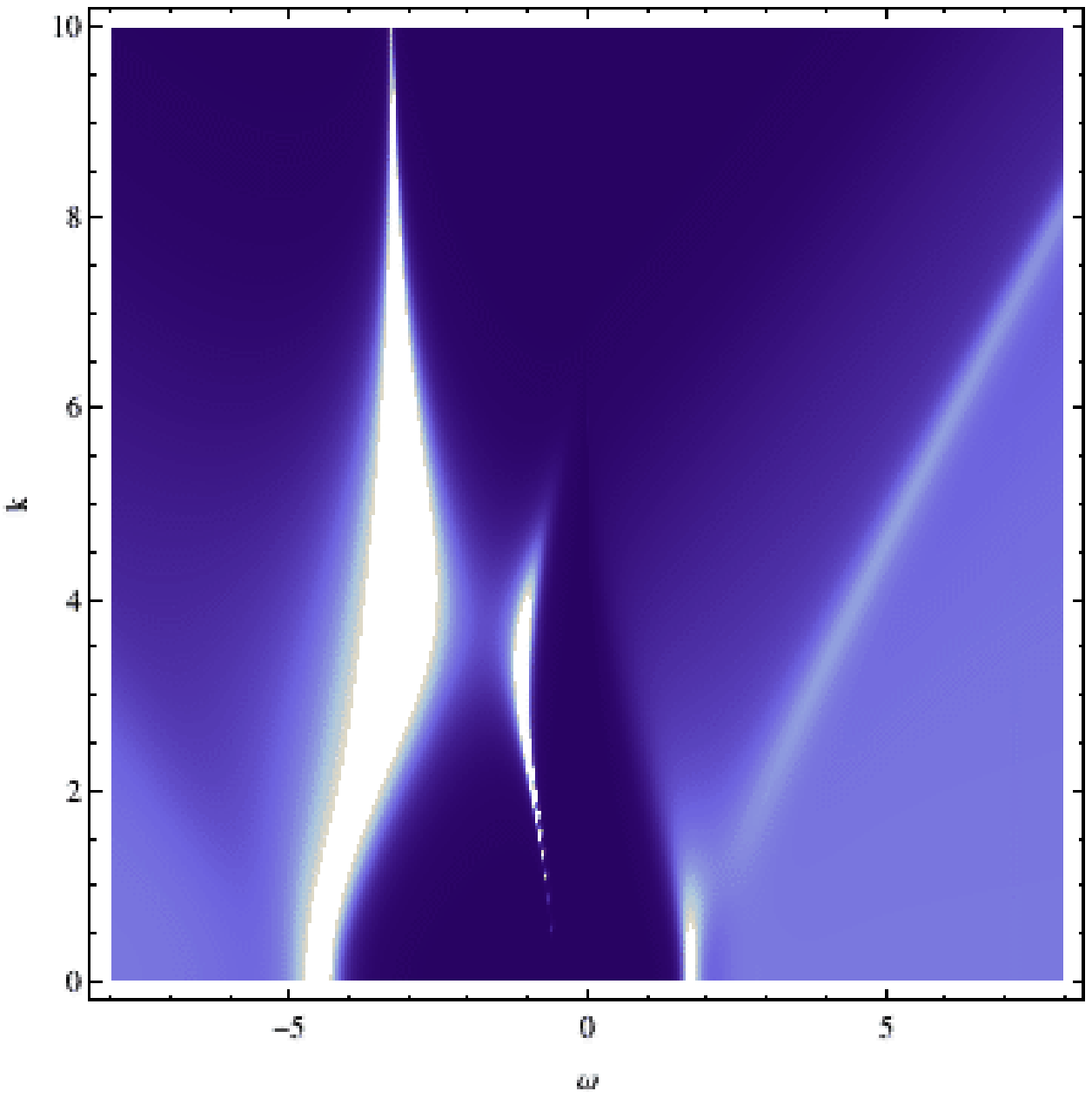}
       \caption{The 3d and density plots of spectral function for the case of $q=2$ and $p=4$, where the gap bends towards $\omega=0$ with the gap energy $\omega=-0.56$ for the zero momentum mode.     }
       \label{p4q2}
\end{figure}
\begin{figure}
\centering
\includegraphics[width=.48\textwidth]{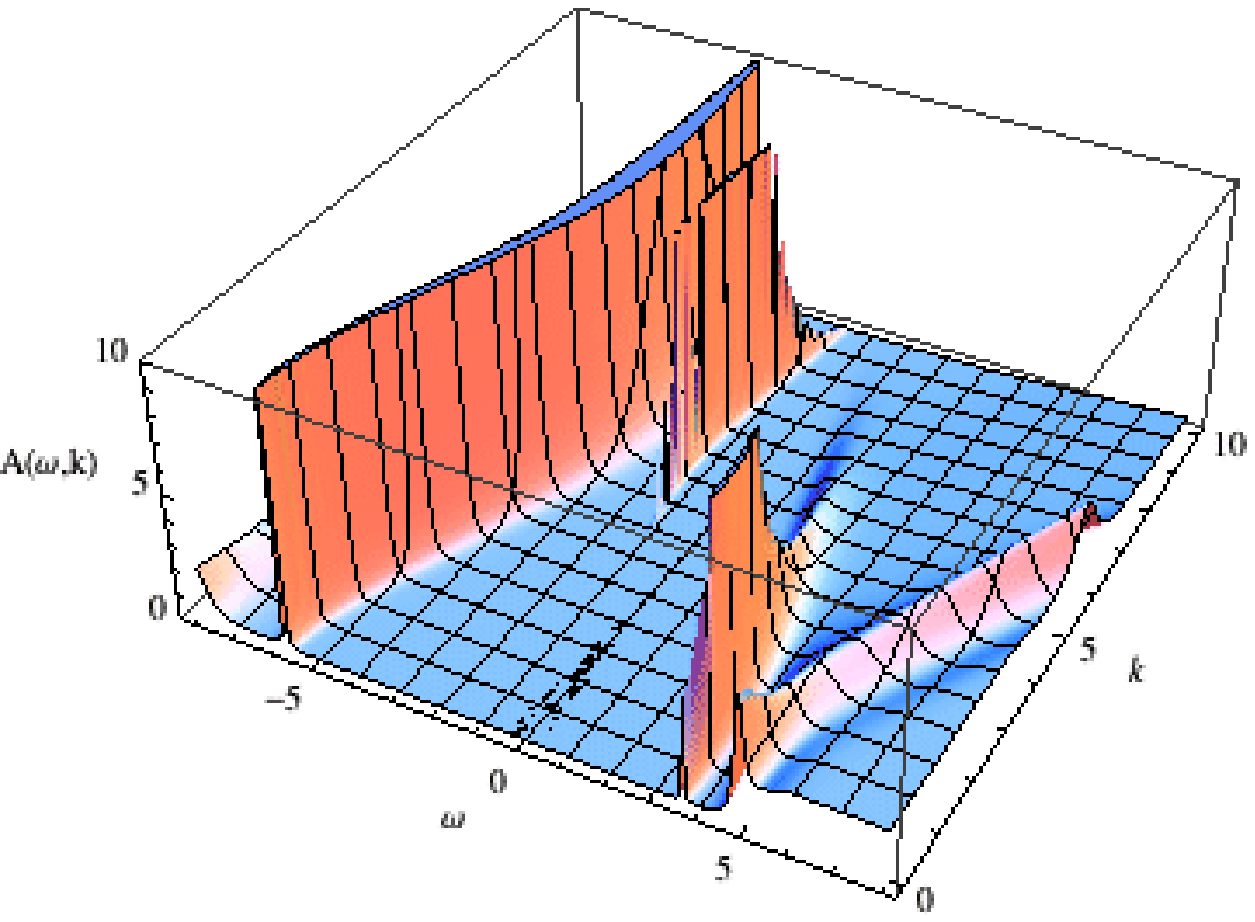}
      \includegraphics[width=.48\textwidth]{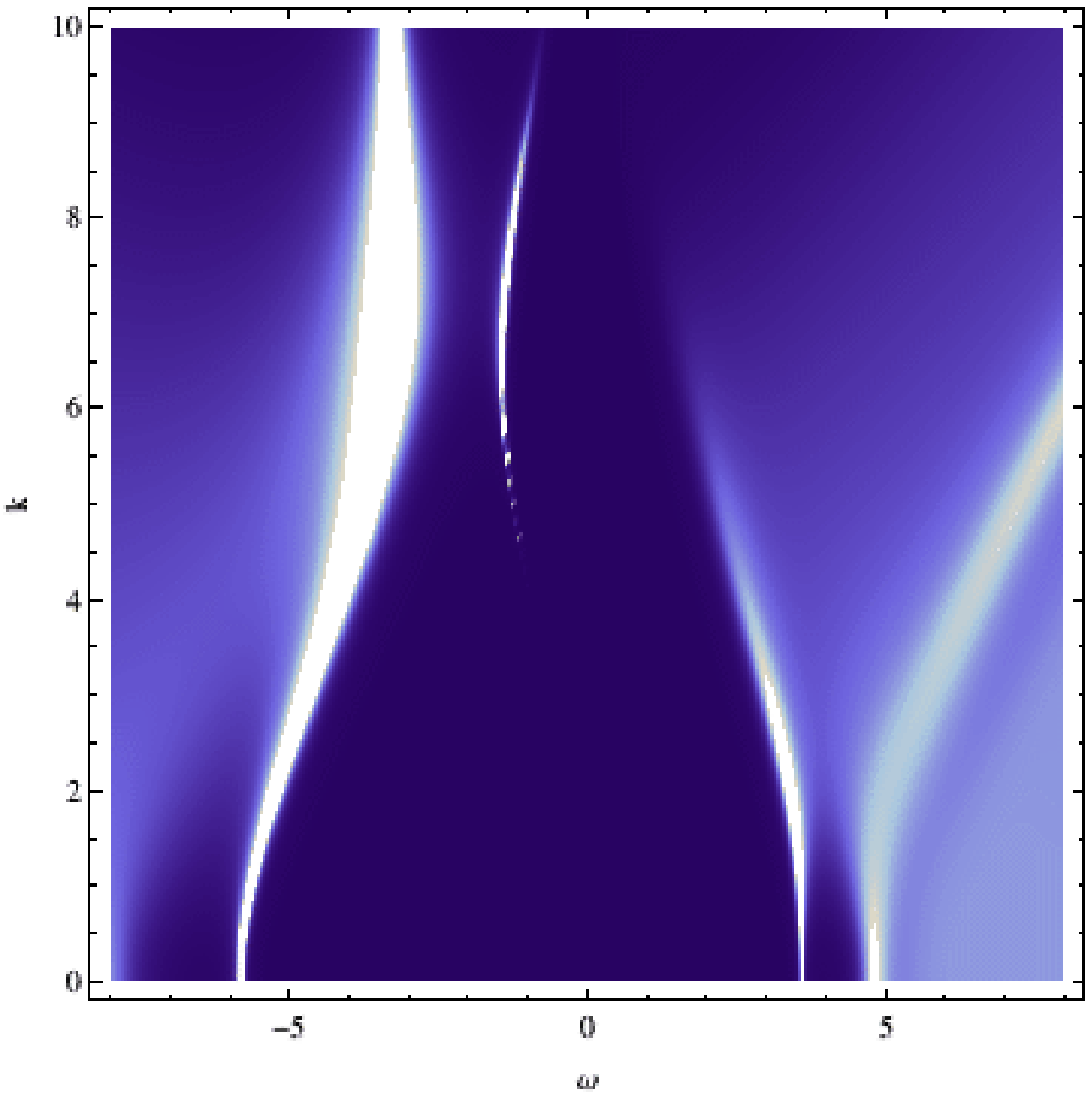}
       \caption{The 3d and density plots of spectral function for the case of $q=2$ and $p=8$,  where the bent gap sprouts a new Fermi surface with the Fermi momentum $k_F=1.20996$.     }
       \label{p8q2}
\end{figure}
Now let us move on to the case of $q=2$. As shown in Figure
\ref{p0q2}, \ref{p2q2}, \ref{p4q2}, and \ref{p8q2}, some new
features come in, namely, the original Fermi surface disappears,
accompanied by a gap opening up at negative $\omega$ as one
increases the dipole coupling $p$ from zero.  But as $p$ is
increased further, the gap gets bent gradually towards $\omega=0$.
Finally the gap closes up, followed by the sprouting of a new Fermi
surface. This suggests that there is some kind of  competition
between the charge $q$ and the bulk dipole coupling $p$. Each of
them prefers to create their own Fermi surfaces. The gap exists at
the parameter place where they counterbalance each other. Such a
pattern can be checked to persist for other larger charge cases.
Finally, we would like to emphasize that such a rebirth of Fermi
surface from the generated gap like a phoenix is never occurring at
the relativistic fixed points, where instead the gap is generically
enlarged as the bulk dipole coupling is increased\cite{ELLP}.
\section{Conclusions}
We have worked with one of the recently discovered non-relativistic
fermionic fixed points and investigated how the corresponding
spectral function is modified by the bulk dipole term numerically.
As a result, although the infinite flat band is robust against the
bulk dipole coupling as well as chemical potential, the bulk dipole
coupling modifies the flat band in particular at the momenta lower
than the value of bulk dipole coupling, bending the band towards
more negative frequencies. More remarkably, it is shown that a new
Fermi surface emerges from the gradually disappearing gap as the
bulk dipole coupling starts to outweigh the charge parameter, which
is totally different from what is happening at the relativistic
fixed points. It is definitely worthwhile to identify such features
by scanning the realistic condensed matter systems and see what kind
of role the bulk dipole coupling mimics there.

 We conclude with various directions worthy of further investigation in the near future.
 First, although for the cases of $m\neq0$ the
retarded Green function is expected to exhibit the qualitatively
similar behaviors as the massless story we have told above, it is
interesting to investigate how some specific features of retarded
Green function like the location of Fermi momentum depend
quantitatively on the mass $m$. Second, it is intriguing to heat
up our system to a finite temperature by considering the charged AdS
black hole away from the extremality to see which features are
smoothed out by such a finite temperature while which features can
persist against the temperature. Last but not least, as is well
known, the low energy behaviors around the Fermi surface are
controlled by the $AdS_2\times R^2$ region near the horizon for the
relativistic fixed points. On the other hand, only in how to massage
the boundary data does the difference lie between the
non-relativistic and relativistic fixed points. The bulk dynamics is
the same for all of these fixed points. So it is expected that at
the non-relativistic fixed points the low energy behaviors around the
Fermi surface are also determined by the emergent IR CFT associated
with the $AdS_2\times R^2$ region somehow. In particular, it is highly
desirable to have an analytic understanding of the low energy
behaviors around the Fermi surface at the non-relativistic fixed
points by performing the rigmarole of matching calculation. We expect to explore these issues elsewhere.
\section*{Acknowledgements}
We would like to thank David Tong for his insightful and illuminating correspondence during the whole course of this project. WJL is grateful to Sijie Gao for his everlasting encouragement and stimulating discussions. He is also indebted to Xiaomei Kuang
for her help to run part of our Mathematica programs in her personal
computer and Jianpin Wu for his sharing the experience in identifying the Fermi surface numerically. HZ is grateful to John McGreevy during PASCOS 2011 in Cambridge and Robert Leigh during the 41th Summer
Institute in Paris for their valuable explanations of their independent
works, which are relevant to our project. HZ is also indebted to
Rene Meyer and Zhi Wang for helpful discussions. In addition, he appreciates Ioannis Iatrakis and Matti Jarvinen for their very help with our Mathematica codes. He also acknowledges the Simons Summer Workshop in Mathematics and Physics 2011 in Stony Brook for the fantastic hospitality during the initial stage of this project. WJL was supported in part by
the NSFC under grant No.10605006 together with 10975016 and by the
Fundamental Research Funds for the Central Universities.
HZ was partially supported by a European Union grant FP7-REGPOT-2008-1-CreteHEPCosmo-228644. This research was also supported in part by the Project of Knowledge Innovation Program of Chinese Academy of Sciences, Grant No.KJCX2.YW.W10.

\end{document}